\documentclass[prl,twocolumn,superscriptaddress,amscd,amsmath,amssymb,verbatim]{revtex4}

\usepackage{graphicx,color}
\usepackage{textcomp}
\usepackage[OT1,T1]{fontenc}
\usepackage[squaren, Gray, cdot]{SIunits}
\usepackage{amsmath}
\usepackage{nicefrac}
\usepackage{natbib}
\usepackage{revsymb}

\newcommand{\GIO}{Gd$_2$Ir$_2$O$_7$}

\begin{document}

\title{Gd pyrochlore under a staggered molecular field in \GIO}

\author{E. Lefran\c cois}
\altaffiliation[]{current address: Max Planck Institute for Solid State Research, Stuttgart, Germany}
\affiliation{Institut Laue Langevin, CS 20156, 38042 Grenoble, France}
\affiliation{Institut N\'eel, CNRS \& Univ. Grenoble Alpes, F-38042 Grenoble, France}
\author{L. Mangin-Thro}
\affiliation{Institut Laue Langevin, CS 20156, 38042 Grenoble, France}
\author{E. Lhotel}
\affiliation{Institut N\'eel, CNRS \& Univ. Grenoble Alpes, F-38042 Grenoble, France}
\author{J. Robert}
\affiliation{Institut N\'eel, CNRS \& Univ. Grenoble Alpes, F-38042 Grenoble, France}
\author{S. Petit}
\affiliation{Laboratoire L\'eon Brillouin, CEA, CNRS, Univ. Paris-Saclay , F-91191 Gif-sur-Yvette, France}
\author{V. Cathelin}
\affiliation{Institut N\'eel, CNRS \& Univ. Grenoble Alpes, F-38042 Grenoble, France}
\author{H. E. Fischer}
\affiliation{Institut Laue Langevin, CS 20156, 38042 Grenoble, France}
\author{C. V. Colin}
\affiliation{Institut N\'eel, CNRS \& Univ. Grenoble Alpes, F-38042 Grenoble, France}
\author{F. Damay}
\affiliation{Laboratoire L\'eon Brillouin, CEA, CNRS, Univ. Paris-Saclay , F-91191 Gif-sur-Yvette, France}
\author{J. Ollivier}
\affiliation{Institut Laue Langevin, CS 20156, 38042 Grenoble, France}
\author{P. Lejay}
\affiliation{Institut N\'eel, CNRS \& Univ. Grenoble Alpes, F-38042 Grenoble, France}
\author{L. C. Chapon}
\affiliation{Diamond Light Source Ltd., Harwell Science and Innovation Campus, Didcot, United Kingdom}
\affiliation{Institut Laue Langevin, CS 20156, 38042 Grenoble, France}
\author{V. Simonet}
\affiliation{Institut N\'eel, CNRS \& Univ. Grenoble Alpes, F-38042 Grenoble, France}
\author{R. Ballou}
\affiliation{Institut N\'eel, CNRS \& Univ. Grenoble Alpes, F-38042 Grenoble, France}

\begin{abstract}
The influence of a staggered molecular field in frustrated rare-earth pyrochlores, produced via the magnetic iridium occupying the transition metal site, can generate exotic ground states, such as the fragmentation of the magnetization in the Ho compound. At variance with the Ising Ho$^{3+}$ moment, we focus on the behavior of the quasi isotropic magnetic moment of the Gd$^{3+}$ ion at the rare-earth site. By means of macroscopic measurements and neutron scattering, we find a complex situation where different components of the magnetic moment contribute to two antiferromagnetic non-collinear arrangements: a high temperature all in -- all out order induced by the Ir molecular field, and Palmer and Chalker correlations that tend to order at much lower temperatures. This is enabled by the anisotropic nature of the Gd-Gd interactions and requires a weak easy-plane anisotropy of the Gd$^{3+}$ moment due to the mixing of the ground state with multiplets of higher spectral terms.
\end{abstract}

\maketitle

\begin{figure}[h!]
	\centering
	\includegraphics[width=8.8cm]{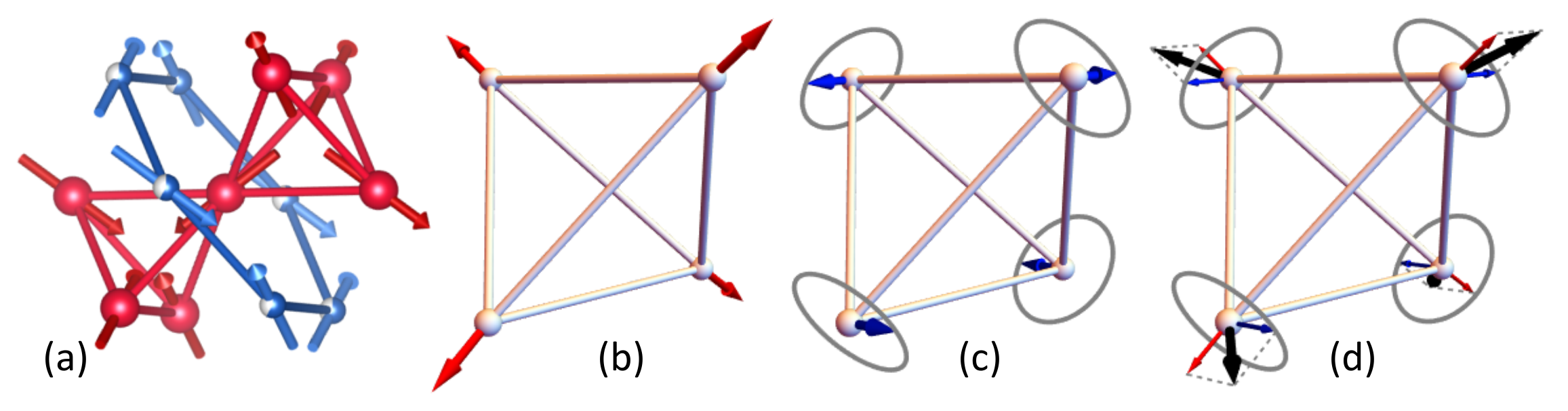}
	\caption{(a) Sketch of the AIAO ordering produced on the rare-earth (red) by the iridium AIAO order (blue). Scheme of different spin configurations: (b) AIAO along the $\langle 111 \rangle$ directions; (c) PC in the planes perpendicular to the $\langle 111 \rangle$ directions \cite{footnote}; (d) Total magnetic moments (black) obtained from the coexistence of AIAO (red) and PC (blue) components. 
}
	\label{picture}
\end{figure}

Pyrochlore compounds, of formula $R_2M_2$O$_7$ with $R^{3+}$ a rare-earth ion and $M^{4+}$ a metal ion, exhibit rich physics due to magnetic frustration \cite{Gardner10}. Depending on the rare-earth ion, that occupies a pyrochlore lattice made of corner-sharing tetrahedra, different magnetocrystalline anisotropies and magnetic interactions are present, giving a great variety of magnetic behaviors.  The presence of a magnetic metal ion, like Ir$^{4+}$, expands further the possible phases compared to the case of non-magnetic $M$. For most $R_2$Ir$_2$O$_7$ compounds, the Ir sublattice orders at rather high temperature (above 100~K for rare-earths heavier than Nd) in the so-called all in -- all out (AIAO) configuration where all the magnetic moments point alternatively inward / outward each tetrahedron \cite{Matsuhira11,Sagayama13,Lefrancois15,Guo16}. This produces a staggered molecular field on the rare-earth sublattice along the local $\langle 111 \rangle$ directions of the cubic unit cell, which can polarize the rare-earth magnetic moments (see Fig. \ref{picture}(a)). This is what happens for rare-earths with an easy-axis anisotropy along the $\langle 111 \rangle$ directions, so that the rare-earth sublattice orders magnetically in the AIAO configuration (see Fig. \ref{picture}(b)), as in the Tb and Ho compounds \cite{Lefrancois15,Lefrancois17}. At lower temperature, typically in the kelvin range, the $R$-$R$ interactions become prominent. They actually slightly modify the Tb magnetic order \cite{Guo17} and have drastic effects on the Ho properties. In the latter case, the effective ferromagnetic interactions between the rare-earth ions compete with the staggered molecular field. This leads to the fragmentation of the magnetization, in which half of the Ho magnetic moment participates to an AIAO ordered state and the other half to a fluctuating Coulomb phase \cite{Lefrancois17}. A very different behavior is observed in the case of Er which has an easy-plane anisotropy, i.e. the Er magnetic moments are perpendicular to the Ir molecular field. No order is detected by magnetometry down to 70 mK and the nature of the ground state remains unknown \cite{Lefrancois15}.

Among this family, the case of the Gd ion is singular. As a first approximation, the Gd$^{3+}$ zero orbital momentum (4f$^7$, $S$=7/2, $L$=0) should lead to isotropic magnetic properties. 
Gd based pyrochlore compounds with non magnetic $M$ ions attracted interest as possible realizations of the Heisenberg pyrochlore antiferromagnet where a spin liquid ground state was predicted \cite{Reimers92, Moessner98}. Real materials were however found to show different behaviors. For $M=$\ Sn, an ordered state with antiferromagnetic pairs, perpendicular to each other, on each tetrahedron, was identified below 1.4 K \cite{Wills06} (see Fig. \ref{picture}(c)). This Palmer-Chalker (PC) state had been predicted for a pyrochlore Heisenberg antiferromagnet in the presence of dipolar interactions \cite{Palmer00}. In the Ti case, two phase transitions are observed, resulting in a partially disordered magnetic state, whose nature is still debated.
These peculiar properties were ascribed to additional ingredients such as further neighbor interactions or additional anisotropies \cite{Champion01,Stewart04,Cepas04,Paddison15,Javanparast15}. 

In this letter, we investigate the magnetic behavior of a Gd pyrochlore with Ir on the metal ion site. Given the strength of the Ir molecular field, and the absence of strong single ion anisotropy, one would naturally expect the Gd magnetic moments to align along the AIAO Ir molecular field. 
We show however that, in addition to the AIAO ordering, a PC component develops at low temperature due to anisotropic Gd-Gd interactions and weak easy-plane single-ion Gd$^{3+}$ anisotropy, both competing with the Ir molecular field. Our analysis emphasizes the role of the Gd anisotropy resulting from mixing of the ground state with multiplets of higher spectral terms, which is often overlooked in studies of Gd compounds. 

\begin{figure}[h]
	\centering
	\includegraphics[width=8.8cm]{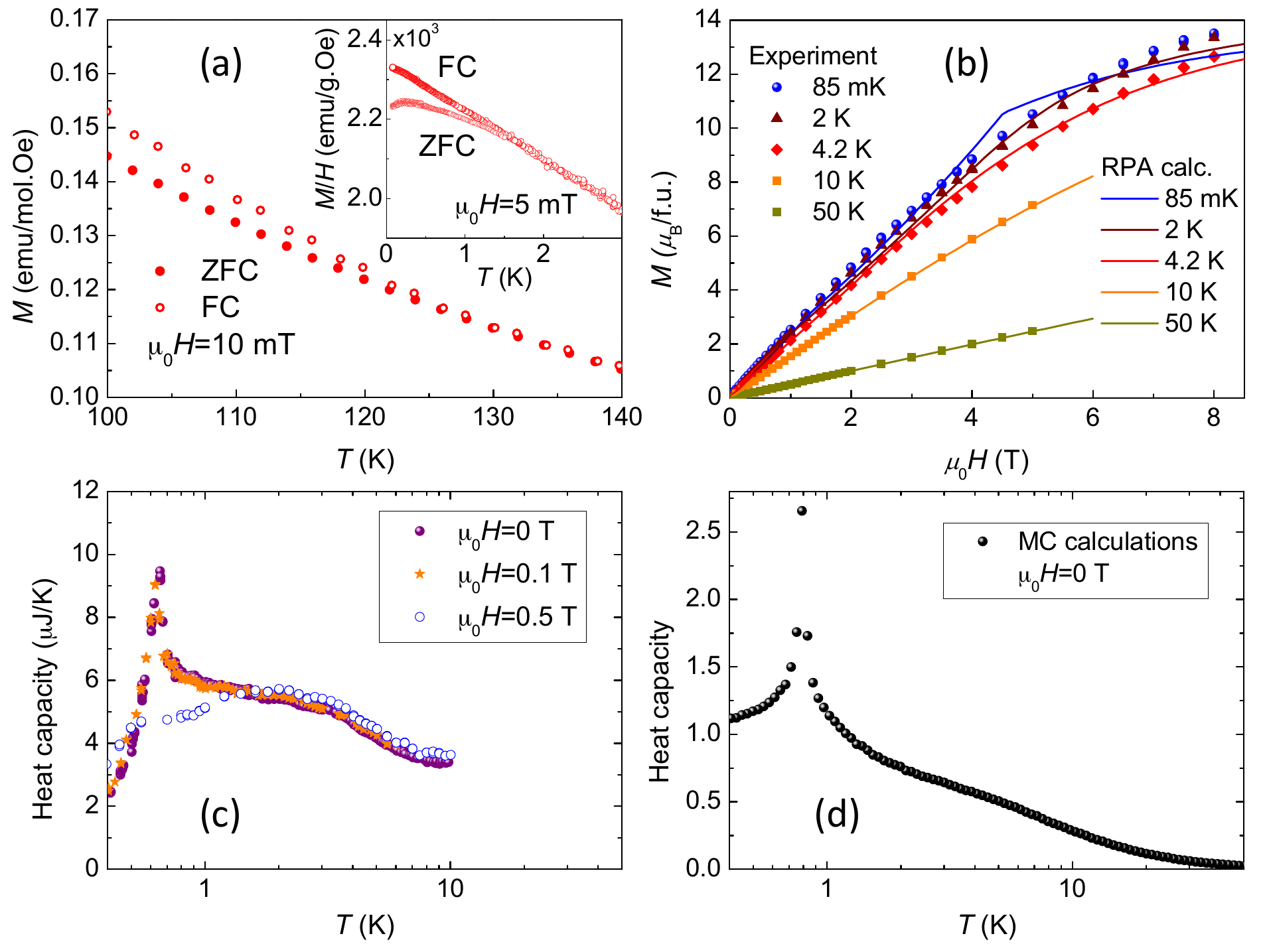}
	\caption{(a) High temperature ZFC-FC magnetization of Gd$_2$Ir$_2$O$_7$ for $\mu_0 H = 10$ mT. Inset:  ZFC-FC magnetization between 0.08 and 4 K for $\mu_0 H = 5$ mT. (b) Magnetization curves versus field at different temperatures and powder averaged $M(H)$ calculated using RPA. 
Specific heat versus temperature in a semilogarithmic scale: (c) measurements at various magnetic fields and (d) zero-field MC calculations.} 
	\label{Mag}
\end{figure}


Pyrochlore iridates crystallize in the Fd$\bar 3$m cubic space group, with the Gd and Ir occupying the 16d and 16c Wyckoff positions, respectively \cite{Gardner10}. \GIO\ polycrystalline samples with natural Gd and isotopic $^{160}$Gd were synthesized by a mineralization process. The starting reagents Gd$_2$O$_3$ and IrO$_2$ were mixed together with a small amount of potassium fluoride flux and pressed into a pellet before undergoing a heat treatment. 
The phase purity and structural quality were checked by x-ray diffraction for both samples. Their lattice parameter and the $x$ coordinate of the 48f O were found at room temperature equal to 10.295(1)/10.277(1)~\AA\ and 0.333(1)/0.343(1) for the natural and isotopic samples respectively. 


The sample with natural Gd was used in the diffraction experiment \cite{doi_D4cD1B} performed on the D4c hot neutron diffractometer of the ILL \cite{Fischer02}, with an incoming wavelength $\lambda=0.50$ \AA. The $^{160}$Gd sample was used in the diffraction measurements at ILL on the thermal neutron diffractometer D1B with $\lambda$=2.52 \AA\  \cite{doi_D4cD1B} and on the polarized neutron diffuse scattering spectrometer D7 with $\lambda$=3.12~\AA\ \cite{doi_D7}. On D7, the XYZ neutron polarization analysis method allowed to extract the magnetic signal \cite{Scharpf93}. Magnetic excitations were measured on the same sample on the IN6 time-of-flight spectrometer at ILL with $\lambda$=5.1~\AA\ \cite{doi_IN6}. The high temperature magnetometry (2-300 K) was performed on both samples with similar results using a Quantum Design MPMS SQUID magnetometer. The low temperature magnetometry (0.08-4.2~K) was performed on the natural Gd sample using a purpose-built SQUID magnetometer equipped with a dilution refrigerator \cite{Paulsen01}. The specific heat of a Gd$_2$Ir$_2$O$_7$ pellet was measured with a Quantum Design PPMS relaxation-time calorimeter from 0.4 to 10 K using a $^3$He insert. Calculations were performed using an hybrid Monte Carlo (MC) method with a single-spin-flip Metropolis algorithm. The MC method is combined to an integration of the nonlinear coupled equations of motion for spin dynamics to obtain the dynamical scattering function $S(\mathbf{Q},\omega)$ \cite{Taillefumier14}. Complementary calculations were carried out in the Random Phase Approximation (RPA) \cite{Petit2012}, which describes better the ordered phase properties but overestimates the transition temperatures.


\begin{figure}[h]
	\centering
	\includegraphics[width=8.7cm]{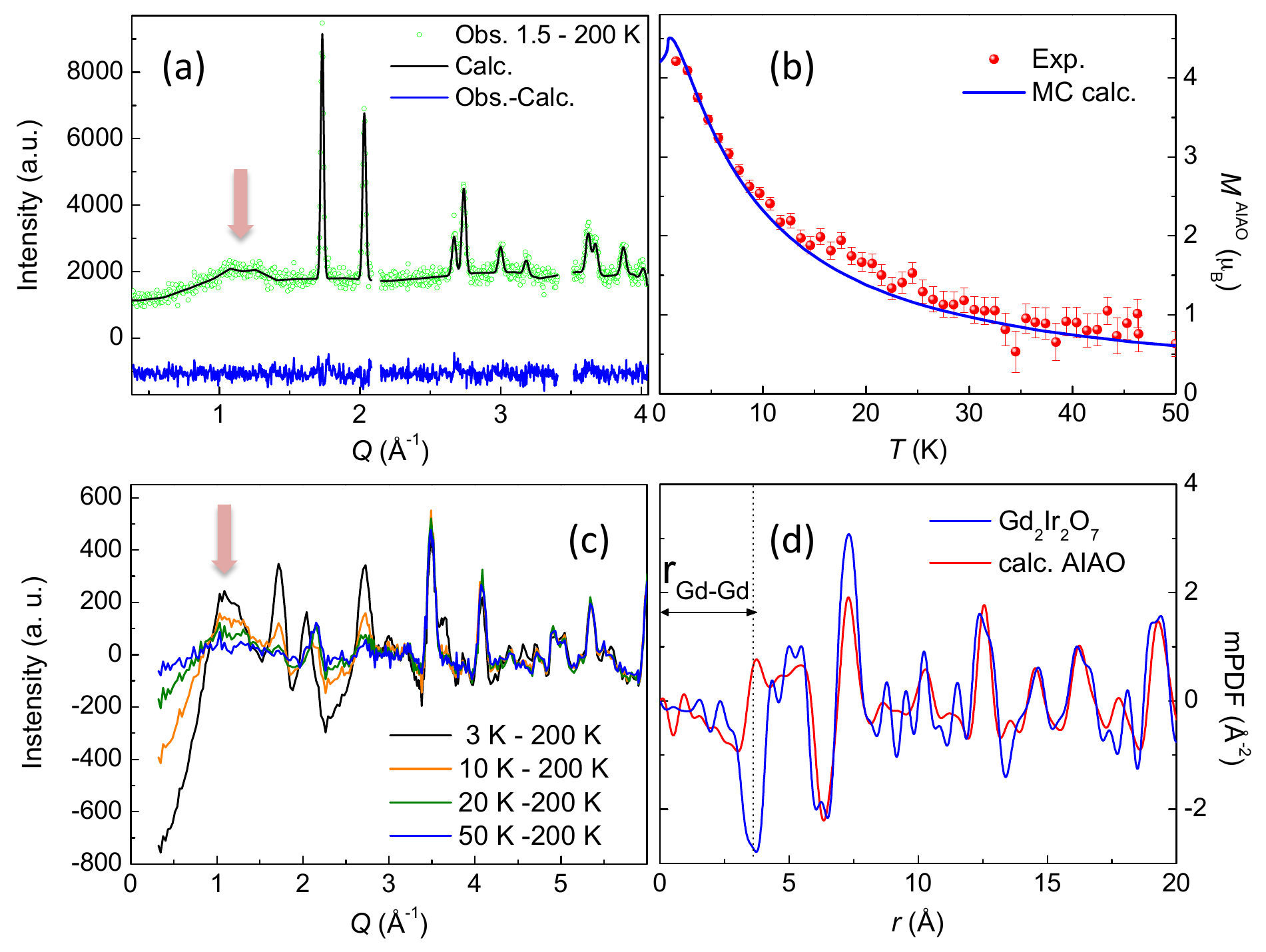}
	\caption{Neutron diffraction results on Gd$_2$Ir$_2$O$_7$. 
	(a) Difference between the D1B diffractograms recorded at 200 and 1.5~K. The Bragg peaks are refined with an AIAO magnetic order on 
the Gd and Ir sites. (b) Temperature dependence of the refined AIAO ordered Gd$^{3+}$ magnetic moment, compared to MC calculations. (c) Difference between the D4c diffractograms at low temperatures (3, 10, 20 and 50 K) and 200 K displayed only up to 6 \AA$^{-1}$. (d) mPDF obtained by Fourier transforming the $3-50$~K D4c diffractogram in the $0-10$~\AA$^{-1}$ $Q$-range (blue) compared to mPDF similarly obtained from a calculated AIAO order diffractogram corrected from the square of the magnetic form factor (red).} 
	\label{Diff}
\end{figure}


Figure \ref{Mag}(a) shows the magnetization versus temperature measured following a zero field cooled - field cooled procedure (ZFC-FC). A separation of the two curves below T$_{\rm Ir}$=120 K indicates the ordering of the Ir$^{4+}$ sublattice, coinciding with the metal-insulator transition reported in these pyrochlore iridates \cite{Matsuhira11}. To determine the magnetic structure below T$_{\rm Ir}$, neutron diffraction experiments were performed (see Fig.~\ref{Diff}). Magnetic Bragg peaks are observed in the diffractograms difference between 1.5 and 200 K. They can be indexed with a ${\bf k=0}$ propagation vector and their intensities are well accounted for by an AIAO magnetic arrangement of both the Gd$^{3+}$ and the Ir$^{4+}$.
Although weak compared to the Gd$^{3+}$ ones, the inclusion of the Ir$^{4+}$ magnetic moments, refined to 0.30(3)~$\mu_{\rm B}$ below 50 K, significantly improves the goodness of the Rietveld fit performed using the Fullprof software \cite{Rodriguez93}. The Gd$^{3+}$ ordered magnetic moment shows a strong increase below 50 K with a maximum value of 4.35(4) $\mu_{\rm B}$ at 1.5 K (See Fig.~\ref{Diff}(b)). This temperature dependence is characteristic of an induced order of the Gd magnetic moment by the AIAO Ir molecular field, through an Ir-Gd magnetic coupling. 

However, an additional diffuse scattering appears below 50 K as a broad bump with a maximum around 1.15~\AA$^{-1}$ (see arrows in Figs \ref{Diff}(a) and (c)). The nature of this bump can be understood by performing, on the D4c magnetic diffractogram, a mPDF (magnetic pair distribution function) analysis which provides equal sensitivity for short- and long-range orders visualized in real space \cite{Frandsen14}. The obtained mPDF is compared in Fig. \ref{Diff}(d) to the calculated mPDF for a pure AIAO magnetic long-range order. A clear difference is observed around 3.6~\AA, the distance between Gd first neighbors, where a negative peak is present in the Gd$_2$Ir$_2$O$_7$ data. Such a negative peak is associated with antiferromagnetic correlations of the spin components in the direction perpendicular to the pair bond \cite{Frandsen14} and is ascribed to the Gd-Gd first neighbor interactions. 

The onset of these Gd spin-spin correlations, different from the AIAO ones, might explain the irreversibility at about 1 K in the low temperature ZFC-FC magnetization (see inset of Fig. \ref{Mag}(a)) and the broad signal around 2 K in the specific heat (see Fig.~\ref{Mag}(c)). These correlations finally lead to a magnetic phase transition at about 650 mK, which manifests as a sharp peak in specific heat, and is smeared out for a magnetic field of 0.5~T. 

\begin{figure}[h]
	\centering
	\includegraphics[width=8.5cm]{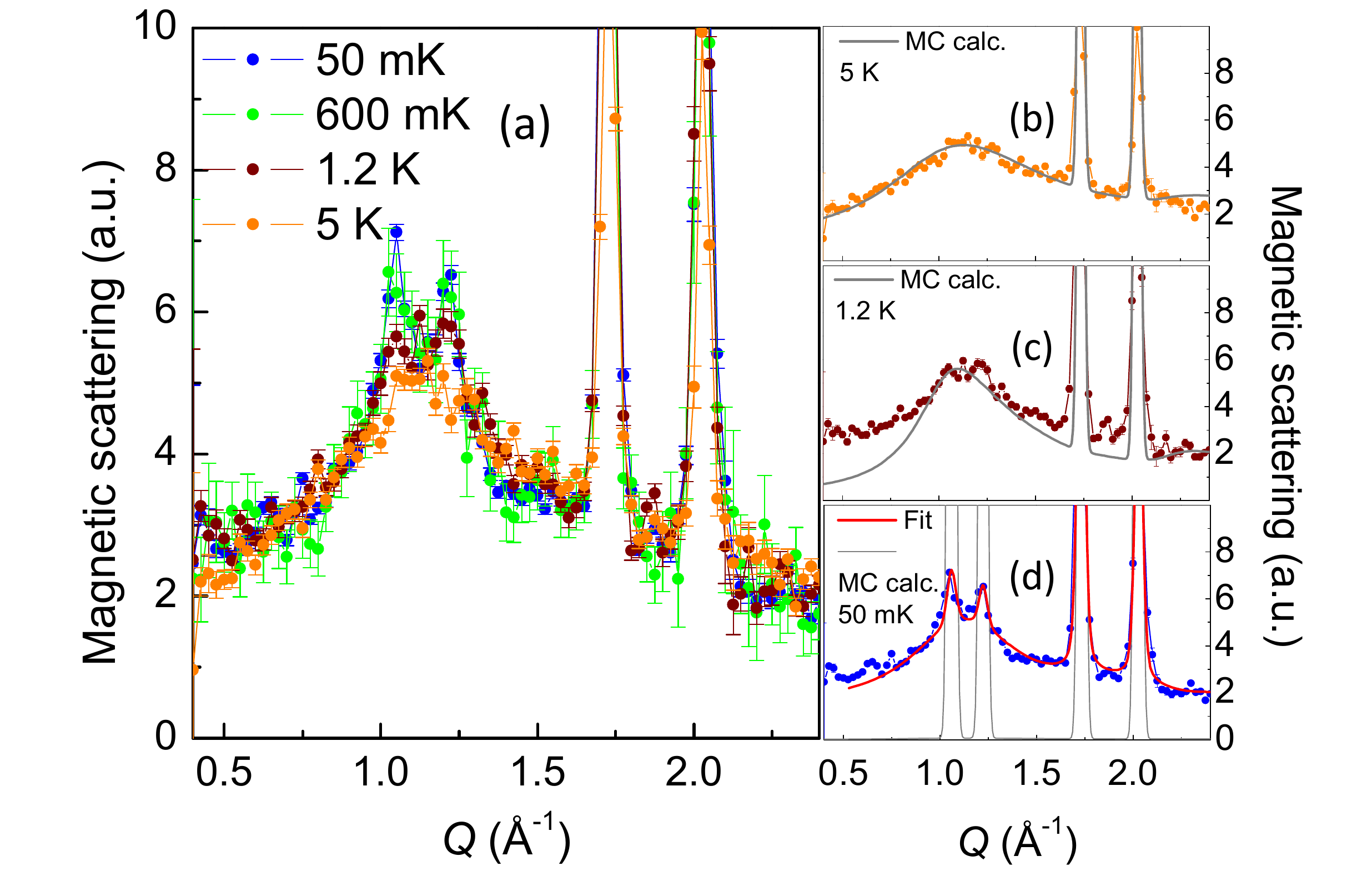}
	\caption{ (a-d) Magnetic scattering of \GIO\ from D7 polarized neutron measurements at various temperatures. 
	(b-d) Grey lines: MC simulations of equal-time scattering functions using the instrumental $Q$-resolution. 
	(d) Red line: Fullprof refinement of coexisting AIAO and PC components of the magnetic moment.}
	\label{Pol}
\end{figure}

The coexistence of AIAO Bragg peaks with diffuse scattering at 5 K was confirmed by polarized neutron diffraction (see Fig. \ref{Pol}). Below 1.2 K, the signal evolves towards a pattern typical of a PC magnetic configuration (see Figs. \ref{Pol}(c)-(d)), characterized by two peaks at 1.05 and 1.22 \AA$^{-1}$ \cite{Petit17}. Consistent with the mPDF analysis, this demonstrates that the Gd magnetic moments eventually have, at low temperature, a PC component perpendicular to the <111> directions. 
We estimate, from a Fullprof refinement shown in Fig. \ref{Pol}(d), the AIAO and PC components to $M_{\parallel}$=4.40(4) and $M_{\perp}$=5.50(4) $\mu_{\rm B}$, respectively, at 1.2~K. This yields a total magnetic moment of $7.04(13)~\mu_{\rm B}$ in agreement with the expected value for Gd$^{3+}$.
In spite of the well-defined specific heat peak, this PC order is not totally long-range since only 1.2 $\mu_{\rm B}$ are in the resolution-limited PC Bragg peaks at 50 mK while the remaining signal is more diffuse (correlation length estimated to 7~\AA). 

\begin{figure}[h]
	\centering
	\includegraphics[width=8.5cm]{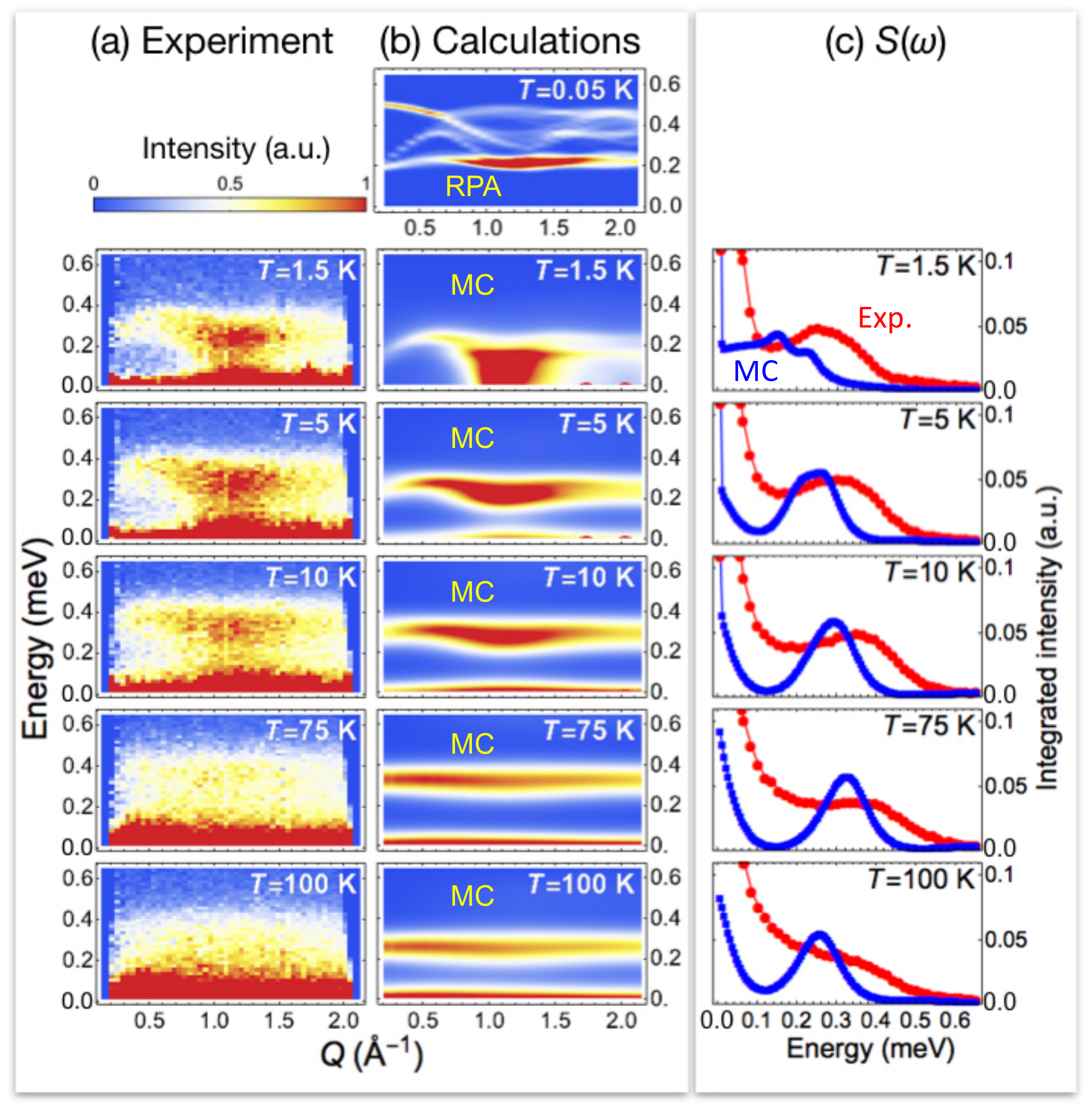}
	\caption{ (a) Temperature dependence of the inelastic neutron scattering of Gd$_2$Ir$_2$O$_7$ measured on IN6 compared to (b) MC calculations from 100 to 1.5 K and RPA calculations in the ordered phase at 50 mK. (c) Corresponding measured (red) and calculated (blue) $Q$-integrated $S(\omega)$.}
	\label{INS}
\end{figure}

Inelastic neutron scattering experiments were performed to probe the excitation spectrum associated to this peculiar magnetic state (see Fig. \ref{INS}). The only visible feature is a quasielastic signal at high temperature that transforms at 100 K into a non-dispersive magnetic excitation peaked around 0.3 meV.  In a first approximation, the Gd$^{3+}$ ion displays no orbital component and is therefore insensitive to the local crystalline electric field (CEF). The observed inelastic signal is then explained by the splitting of the spin ground octet $^{8}S_{7/2}$ of the Gd$^{3+}$ by the Ir molecular field. Due to Gd-Gd interactions, this excitation becomes modulated at lower temperature with a minimum at around 1.15 \AA$^{-1}$, the momentum transfer where the PC diffuse scattering has maximum intensity. 


Actually, a small admixture with multiplets of non zero orbital moments is possible. This can confer to the Gd$^{3+}$ magnetic moments a very weak easy-plane anisotropy, as reported for the Sn and Ti isostructural compounds \cite{Glazkov05, Glazkov07}, and can also result in anisotropic exchange. Although weak, this orbital contribution is a key ingredient for the interpretation of the peculiar magnetic properties of Gd$_2$Ir$_2$O$_7$ as discussed below.

In order to reproduce the experimental results, we considered the Hamiltonian:
\begin{equation}
\begin{aligned}
\label{eqH}
{\cal H} = &J\sum_{\langle i,j\rangle}{{\bf S}_i \cdot {\bf S}_j } +D_{pd}\sum_{\langle i,j\rangle}{\left[{\bf S}_i\cdot {\bf S}_j-\frac{3({\bf S}_i\cdot {\bf r}_{ij})({\bf S}_j\cdot {\bf r}_{ij})}{r^2}\right]} \\ 
&- g \mu_0{\mu_{\rm B}} H^{loc} \sum_i { \mathbf{z}_i \cdot \mathbf{S}_i} + \sum_{i} B_2^0 (O_2^0)_i
\end{aligned}
\end{equation} 
where ${\bf S}_i$ is the $i^{th}$ Gd$^{3+}$ magnetic moment, $J>0$ the Gd-Gd first neighbor antiferromagnetic isotropic exchange interaction and 
$D_{pd}$ defines an anisotropic interaction limited to first neighbors with $r$ their separation distance. This term has the same anisotropy as the first neighbor dipolar interaction, for which $D_{pd}=D_{dip}=(g\mu_0 \mu_{\rm B})^2/r^3$. 
$H^{loc} = 6 J_{zz}^{\mathrm{Gd-Ir}} m_{\rm Ir}/(g\mu_0 \mu_{\rm B})$ is the AIAO staggered magnetic field produced by the Ir magnetic moments $m_{\rm Ir}$ and mediated by the Ir-Gd interactions $J_{zz}^{\mathrm{Gd-Ir}}$, $\mathbf{z}$ being the local $\langle 111\rangle$ axis. The last term is the easy-plane anisotropy with $B_2^0>0$, $(O_2^0)_i = 3 ( \mathbf{z}_i \cdot \mathbf{S}_i)^2-S(S+1)$. 

At $T=0$, considering only the interaction terms, $J$ and $D_{pd}>0$,  leads to PC magnetic ordering \cite{Palmer00}. However, as soon as $H^{loc}\neq 0$, the AIAO order parameter $M_{\parallel}$ rises at the expense of the PC one $M_{\perp}$. With the full model $-$ including the molecular field and single-ion anisotropy $-$ and following Ref. \citenum{Yan17}, we obtain the ground state energy per spin for a single tetrahedron: 
 \begin{equation}
\begin{aligned}
E_{GS} &=[(5D_{pd}- J)M^2_{\parallel}+(-{5\over 2}D_{pd}-J)M^2_{\perp}] \\
&- g\mu_0 \mu_{\rm B} H^{loc}M_{\parallel} + 3B_2^0 M^2_{\parallel}
\end{aligned}
\label{eqE}
\end{equation}
This demonstrates that the AIAO and PC order parameters are not coupled. The corresponding magnetic orderings can therefore coexist over a wide range of $J$, $D_{pd}$, $H^{loc}$, and  $B_2^0$ values (see Fig. \ref{picture}(d)). The ratio $M_{\perp}/M_{\parallel}$, independent of $J$, is equal to $\tan \theta$ with: 
\begin{equation}
\theta = \arccos{\left(H^{loc}/H_c\right)}, ~\mbox{and}~H_c=\frac{3S(5D_{pd}+2B_2^0)}{g \mu_0 \mu_{\rm B}}
\label{eq:theta}
\end{equation}
where $H_c$ is the critical field above which the full AIAO ordering is achieved. 

We then searched for the best parameters $J$, $D_{pd}$, $H^{loc}$, and $B_2^0$ reproducing the experiments on \GIO. 
Considering isotropic spins ($B_2^0=0$) and dipolar interactions ($D_{pd}=D_{dip}=0.0519$~K), we were not able to reproduce simultaneously the temperature dependence of $M_{\parallel}$ and the value of $\theta$. 
Hence, the presence of significant single-ion anisotropy and anisotropic interactions beyond the dipolar ones are compulsory to obtain coexisting PC and AIAO spin correlations. 

A final set of parameters in good agreement with all our experimental data is $J=0.23$~K, $D_{pd}=0.11$~K, $\mu_0H^{loc} = 2.9$~T and $B_2^0=0.03$~K, the value of the isotropic interaction $J$ being mainly constrained by the isothermal magnetization curves (see Fig. \ref{Mag}(b)). The $B_2^0$ term is found slightly smaller than those of Ti (0.074~K) and Sn (0.047~K) compounds \cite{Glazkov05,Glazkov07}, while the pseudo-dipolar term $D_{pd}$ is larger than $D_{dip}$. Assuming pure dipolar interactions leads to $B_2^0 >0.17$~K, which is large compared to the Ti and Sn compounds values, and thus supports the presence of enhanced anisotropic interactions of pseudo-dipolar form. Other anisotropic couplings are allowed by symmetry \cite{Curnoe08} but are less relevant in this system. Finally, we checked that the long-range part of the dipolar interaction does not change significantly the results with the chosen parameters. 
 
Using these parameters, the calculated specific heat shows a correlation bump followed by a sharp peak around 800 mK as in the experimental data (see Figs. \ref{Mag}(c)-(d)). The temperature dependence of the AIAO ordered moment component is also reproduced (see Fig.~\ref{Diff}(b)) and correctly reflects the refined proportion of PC and AIAO contributions to the magnetic moment. Calculations reproduce the splitting of the Gd$^{3+}$ spin ground octet and its modulation down to 5 K (see Fig.~\ref{INS}), except for a quasielastic signal present in the measurements at all temperatures.

At lower temperature, discrepancies are nevertheless observed. The model predicts a perfect PC long-range ordering below the transition temperature, whereas the measurements rather feature a distribution of correlation lengths. The $M(H)$ curves calculated with RPA below the mean-field ordering temperature also show anomalies revealing field induced magnetic phases (inflexion points at $\mu_0H\sim$1 and 4.3 T) which are not visible in the experiments (see Fig. \ref{Mag}(b)). The measured excitations are broader and less defined than in the calculations. Especially, spin waves clearly develop in the calculations below 5 K (see Fig. \ref{INS}(b)). They are fully gapped in the ordered regime of RPA calculations as expected in the presence of a molecular field and an ordered PC state \cite{Maestro04, Maestro07, Sosin09}. 


Therefore, while our analysis successfully describes the coexistence of PC and AIAO correlations of the Gd$^{3+}$ spin components, it cannot explain the measured incomplete PC ordering. This discrepancy might be attributed to the presence of defects. In pyrochlore iridates, the substitution of Ir$^{4+}$ by non magnetic Ir$^{3+}$  \cite{Yang17} or magnetic Ir$^{5+}$ \cite{Zhu14} has been proposed, which changes the Ir local field. Oxygen defects can also be present and will considerably modify the exchange paths, which is expected to affect more strongly the PC phase stabilized by Gd-Gd interactions than the AIAO phase. In addition to the fragility of the PC order parameter, the presence of defects could explain the ZFC-FC irreversibilities observed in magnetization measurements. This also seems in line with recent theoretical investigations of the influence of defects in easy-plane pyrochlores which have shown that they can impede PC long-range ordering \cite{Andrade18}. This effect could be enhanced if the system is located in a region where the PC order parameter varies rapidly.
This scenario calls to mind the situation of the pyrochlore oxide Yb$_2$Ti$_2$O$_7$, at the vicinity of a phase diagram boundary \cite{Jaubert15, Robert15}. It displays similar partial ordering with strongly sample dependent properties, ascribed to the presence of defects \cite{Arpino17}. 

In conclusion, we have investigated the role of a staggered AIAO magnetic field competing with the rare-earth exchange interactions in \GIO, a new member of the pyrochlore iridate family.  
The Gd-Gd magnetic interactions developing at low temperature lead to a complex antiferromagnetic pattern, with two AIAO and PC orthogonal components, the PC one being ill-ordered. Our analysis points out the role in this frustrated system of usually neglected terms for Gd$^{3+}$ ions, such as weak single-ion anisotropy and anisotropic interactions, to explain our experimental findings.


\acknowledgments
We acknowledge A. Hadj-Azzem and J. Balay for their help in the compound synthesis, D. Dufeu, E. Eyraud and P. Lachkar for their technical support on the high temperature magnetometers and specific-heat calorimeter, and C. Paulsen for allowing us to use his SQUID dilution magnetometers. We acknowledge A. Wildes for the discussions concerning the polarization analysis on D7. J. R. thanks B. Canals for the joint development of the software used for the calculations of the spin dynamics. E. Lh and V. C. acknowledge financial support from ANR, France, Grant No. ANR-15-CE30-0004.



\begin{thebibliography}{99}

\bibitem{Gardner10} J. S. Gardner, M. J. P. Gingras, J. E. Greedan, Rev. Modern Phys. {\bf 82}, 53 (2010).

\bibitem{Guo16} H. Guo, C. Ritter, and A. C. Komarek, Phys. Rev. B {\bf 94}, 161102(R) (2016). 

\bibitem{Sagayama13} H. Sagayama, D. Uematsu, T. Arima, K. Sugimoto, J. J. Ishikawa, E. O'Farrell, and S. Nakatsuji, Phys. Rev. B {\bf87}, 100403(R) (2013). 

\bibitem{Lefrancois15} E. Lefran\c{c}ois, V. Simonet, R. Ballou, E. Lhotel, A. Hadj-Azzem, S. Kodjikian, P. Lejay, P. Manuel, D. Khalyavin, and L. C. Chapon, Phys. Rev. Lett. {\bf 114}, 247202 (2015).

\bibitem{Matsuhira11} K. Matsuhira, M. Wakeshima, Y. Hinatsu, and S. Takagi, J. Phys. Soc. Jap. {\bf 80}, 094701 (2011). 

\bibitem{Lefrancois17} E. Lefran\c{c}ois, V. Cathelin, E. Lhotel, J. Robert, P. Lejay, C.V. Colin, B. Canals, F. Damay, J. Ollivier, B. F\aa k, L.C. Chapon, R. Ballou and V. Simonet, Nature Communications {\bf 8}, 209 (2017). 

\bibitem{Guo17} H. Guo, C. Ritter, and A. C. Komarek, Phys. Rev. B {\bf 96}, 144415 (2017). 

\bibitem{Reimers92}  J. N. Reimers, Phys. Rev. B {\bf 45}, 7287 (1992).

\bibitem{Moessner98} R. Moessner and J. T. Chalker,  Phys. Rev. Lett. {\bf 80}, 2929 (1998).

\bibitem{Wills06} A. Wills, M. E. Zhitomirsky, B. Canals, J.-P. Sanchez, P. Bonville, P. Dalmas de R\'eotier, and A. Yaouanc, J. Phys.: Condens. Matter {\bf 18}, L37 (2006).

\bibitem{footnote} One over the three possible Palmer-Chalker configurations is shown.

\bibitem{Palmer00} S. E. Palmer and J. T. Chalker, Phys. Rev. B {\bf 62}, 488 (2000).

\bibitem{Champion01} J. D. M. Champion, A. S. Wills, T. Fennell, S. T. Bramwell, J. S. Gardner, and M. A. Green, Phys. Rev. B {\bf 64}, 140407(R) (2001).

\bibitem{Stewart04} J. R. Stewart, G. Ehlers, A. S. Wills, S. T. Bramwell, and J. S. Gardner, J. Phys.: Condens. Matter {\bf 16}, L321 (2004).

\bibitem{Cepas04} O. C\'epas, and B. S. Shastry, Phys. Rev. B {\bf 69}, 184402 (2004).

\bibitem{Javanparast15} B. Javanparast, Z. Hao, M. Enjalran, and M. J. P. Gingras,  Phys. Rev. Lett. {\bf 114}, 130601 (2015).

\bibitem{Paddison15} J. A. M. Paddison, A. B. Cairns, D. D. Khalyavin, P. Manuel, A. Daoud-Aladine, G. Ehlers, O. A. Petrenko, J. S. Gardner, H. D. Zhou, A. L. Goodwin, and J. R. Stewart, arXiv:1506.05045.

\bibitem{doi_D4cD1B} E. Lefran\c{c}ois, {\it et al.} (2015).
Institut Laue-Langevin (ILL). doi:10.5291/ILL-DATA.5-31-2406.

\bibitem{Fischer02} H. E. Fischer, G. J. Cuello, P. Palleau, D. Feltin, A. C. Barnes, Y. S. Badyal and J.M. Simonson, 
{\it Appl.\ Phys. A\/} {\bf 74}, S160--S162 (2002).

\bibitem{doi_D7}  V. Simonet, {\it et al.} (2017). Institut Laue-Langevin (ILL). doi:10.5291/ILL-DATA.5-42-452.

\bibitem{Scharpf93} O. Sch\"arpf and H. Capellmann, Phys. Status Solidi A Appl. Res. {\bf 135}, 359 (1993).

\bibitem{doi_IN6} E. Lefran\c{c}ois, {\it et al.} (2015). Institut Laue-Langevin (ILL). doi:10.5291/ILL-DATA.4-01-1478.

\bibitem{Paulsen01} Paulsen, C. in {\it Introduction to Physical Techniques in Molecular Magnetism: Structural and Macroscopic Techniques - Yesa 1999}, edited by F. Palacio, E. Ressouche, and J. Schweizer (Servicio de Publicaciones de la Universidad de Zaragoza, Zaragoza, 2001), p. 1.

\bibitem{Taillefumier14} M. Taillefumier, J. Robert, C. L. Henley, R. Moessner and B. Canals, Phys. Rev. B {\bf 90}, 064419 (2014).

\bibitem{Petit2012} S. Petit, P. Bonville, I. Mirebeau, H. Mutka, and J. Robert, Phys Rev B {\bf 85}, 054428 (2012).

\bibitem{Rodriguez93} J. Rodriguez-Carvajal, Phys. B (Amsterdam) {\bf 192}, 55 (1993).

\bibitem{Frandsen14} B. A. Frandsen, X. Yang, and S. J. L. Billinge, Acta Cryst. {\bf A70}, 3 (2014).

\bibitem{Petit17} S. Petit, E. Lhotel, F. Damay, P. Boutrouille, A. Forget, and D. Colson, Phys. Rev. Lett. {\bf 119}, 187202 (2017).

\bibitem{Glazkov05} V. N. Glazkov, M. E. Zhitomirsky, A. I. Smirnov, H.-A. Krug von Nidda, A. Loidl, C. Marin, and J.-P. Sanchez, Phys. Rev. B {\bf 72}, 020409(R) (2005).

\bibitem{Glazkov07} V. N. Glazkov, M. E. Zhitomirsky, A. I. Smirnov, C. Marin, and J.-P. Sanchez, A. Forget, D. Colson, and P. Bonville, J. Phys.: Condens. Matter {\bf 19}, 145271 (2007).

\bibitem{Yan17} H. Yan, O. Benton, L. Jaubert, and N. Shannon, Phys. Rev. B {\bf 95}, 094422 (2017).

\bibitem{Curnoe08} S. H. Curnoe, Phys. Rev. B {\bf 78}, 094418 (2008).

\bibitem{Maestro04} A. G. Del Maestro and M. J. P. Gingras, J. Phys. Condens. Matter {\bf 16}, 3339 (2004).
\bibitem{Maestro07} A. Del Maestro and M. J. P. Gingras, Phys. Rev. B {\bf 76}, 064418 (2007).
\bibitem{Sosin09} S. S. Sosin, L. A. Prozorova, P. Bonville, and M. E. Zhitomirsky, Phys. Rev. B {\bf 79}, 014419 (2009).

\bibitem{Yang17} W. C. Yang, W. K. Zhu, H. D. Zhou, L. Ling, E. S. Choi, M. Lee, Y. Losovyj, Chi-Ken Lu, and S. X. Zhang, Phys. Rev. B {\bf 96}, 094437 (2017).

\bibitem{Zhu14} W. K. Zhu, M. Wang, B. Seradjeh, Fengyuan Yang, and S. X. Zhang, Phys. Rev. B {\bf 90}, 054419 (2014).

\bibitem{Andrade18} E. C. Andrade, J. A. Hoyos, S. Rachel, and M. Vojta, Phys. Rev. Lett. {\bf 120}, 097204 (2018).

\bibitem{Jaubert15} L. D. C. Jaubert, O.Benton, J. G. Rau, J. Oitmaa, R. R. P. Singh, N. Shannon, and M. J. P. Gingras, Phys. Rev. Lett. {\bf 115}, 267208 (2015).

\bibitem{Robert15} J. Robert, E. Lhotel, G. Remenyi, S. Sahling, I. Mirebeau, C. Decorse, B. Canals, and S. Petit, Phys. Rev. B {\bf 92}, 064425 (2015).

\bibitem{Arpino17} K. E. Arpino, B. A. Trump, A. O. Scheie,T. M. McQueen, and S. M. Koohpayeh, Phys. Rev. B {\bf 95}, 094407 (2017).

\end{thebibliography}
\end{document}